\documentclass[a4paper]{tufte-book}

\usepackage{comment}
\usepackage{listings}

\hypersetup{colorlinks} 

\title{Disappearing frameworks explained}
\date{v1.0, arXiv edition}
\author[Juho Vepsäläinen]{Juho Vepsäläinen}
\publisher{SurviveJS Oy}

\usepackage{lipsum}

\usepackage{booktabs}

\usepackage{graphicx}
\setkeys{Gin}{width=\linewidth,totalheight=\textheight,keepaspectratio}
\graphicspath{{figures/}}

\usepackage{fancyvrb}
\fvset{fontsize=\normalsize}

\usepackage{xspace}

\newcommand{\monthyear}{%
  \ifcase\month\or January\or February\or March\or April\or May\or June\or
  July\or August\or September\or October\or November\or
  December\fi\space\number\year
}

\makeatletter
\renewcommand{\maketitlepage}{%
\begingroup%
\setlength{\parindent}{0pt}

{\fontsize{24}{24}\selectfont\textit{\@author}\par}

\vspace{1.75in}{\fontsize{28}{54}\selectfont\@title\par}

\vspace{0.5in}{\fontsize{14}{14}\selectfont\textsf{\smallcaps{\@date}}\par}

\vfill{\fontsize{14}{14}\selectfont\textit{\@publisher}\par}

\thispagestyle{empty}
\endgroup
}
\makeatother

\newcommand{\openepigraph}[2]{%
  \begin{fullwidth}
  \sffamily\large
  \begin{doublespace}
  \noindent\allcaps{#1}\\
  \noindent\allcaps{#2}
  \end{doublespace}
  \end{fullwidth}
}

\usepackage{units}

\newcommand{\hlred}[1]{\textcolor{Maroon}{#1}}
\newcommand{\hangleft}[1]{\makebox[0pt][r]{#1}}

\providecommand{\XeLaTeX}{X\lower.5ex\hbox{\kern-0.15em\reflectbox{E}}\kern-0.1em\LaTeX}

\newcommand{\tuftebs}{\symbol{'134}}
\newcommand{\doccmddef}[2][]{%
  \hlred{\texttt{\tuftebs#2}}\label{cmd:#2}%
  \ifthenelse{\isempty{#1}}%
    {
      \index{#2 command@\protect\hangleft{\texttt{\tuftebs}}\texttt{#2}}
    }%
    {
      \index{#2 command@\protect\hangleft{\texttt{\tuftebs}}\texttt{#2} (\texttt{#1} package)}
      \index{#1 package@\texttt{#1} package}\index{packages!#1@\texttt{#1}}
    }%
}
\newcommand{\doccmd}[2][]{%
  \texttt{\tuftebs#2}%
  \ifthenelse{\isempty{#1}}%
    {
      \index{#2 command@\protect\hangleft{\texttt{\tuftebs}}\texttt{#2}}
    }%
    {
      \index{#2 command@\protect\hangleft{\texttt{\tuftebs}}\texttt{#2} (\texttt{#1} package)}
      \index{#1 package@\texttt{#1} package}\index{packages!#1@\texttt{#1}}
    }%
}

\usepackage{makeidx}
\makeindex
\definecolor{lightgray}{rgb}{.9,.9,.9}
\definecolor{darkgray}{rgb}{.4,.4,.4}
\definecolor{purple}{rgb}{0.65, 0.12, 0.82}
\definecolor{editorWhite}{rgb}{1, 1, 1}
\definecolor{editorGray}{rgb}{0.95, 0.95, 0.95}
\definecolor{editorOchre}{rgb}{1, 0.5, 0} 
\definecolor{editorGreen}{rgb}{0, 0.5, 0} 

\lstdefinelanguage{JavaScript}{
    morekeywords={break, case, catch, continue, debugger, default, delete, do, else, false, finally, for, function, if, in, instanceof, new, null, return, switch, this, throw, true, try, typeof, var, void, while, with},
    morecomment=[s]{/*}{*/},
    morecomment=[l]//,
    morestring=[b]",
    morestring=[b]'
}

\lstdefinelanguage{CSS} 
{morekeywords={color,background,margin,padding,margin,padding,font,weight,display,position,top,left,right,bottom,list,style,border,size,white,space,min,width, 	transition}, 
	sensitive=false, 
	morecomment=[l]{//}, 
	morecomment=[s]{/*}{*/}, 
	morestring=[b]", 
}

\lstdefinelanguage{HTML5}{
  language=html,
  sensitive=true,	
  alsoletter={<>=-},	
  morecomment=[s]{<!-}{-->},
  tag=[s],
  otherkeywords={
  >,
	<!DOCTYPE,
  </html, <html, <head, <title, </title, <style, </style, <link, </head, <meta, />,
	</body, <body,
	</div, <div, </div>, 
	</p, <p, </p>,
	</script, <script,
  <canvas, /canvas>, <svg, <rect, <animateTransform, </rect>, </svg>, <video, <source, <iframe, </iframe>, </video>, <image, </image>, <header, </header, <article, </article
  },
  ndkeywords={
  =,
  charset=, src=, id=, width=, height=, style=, type=, rel=, href=,
  fill=, attributeName=, begin=, dur=, from=, to=, poster=, controls=, x=, y=, repeatCount=, xlink:href=,
  margin:, padding:, background-image:, border:, top:, left:, position:, width:, height:, margin-top:, margin-bottom:, font-size:, line-height:,
  transform:, -moz-transform:, -webkit-transform:,
  animation:, -webkit-animation:,
  transition:,  transition-duration:, transition-property:, transition-timing-function:,
  }
}

\begin{document}

\frontmatter

\maketitle

\newpage
\begin{fullwidth}
~\vfill
\thispagestyle{empty}
\setlength{\parindent}{0pt}
\setlength{\parskip}{\baselineskip}
Copyright \copyright\ \the\year\ \thanklessauthor

\par\smallcaps{Published by \thanklesspublisher}

\par\smallcaps{https://tufte-latex.github.io/tufte-latex/}

\par\textit{First printing, \monthyear}
\end{fullwidth}

\tableofcontents

\listoffigures

\cleardoublepage
\chapter*{Introduction}

The version of this short book you are reading is a version that has been tailored for you, the readers of arXiv. In other words, I may expand it into something larger one day, and what you see here can be considered a preview and a brief introduction to the rising topic of disappearing frameworks. If you have requests or ideas on how to develop the content further, please email me at info@survivejs.com.

\noindent\makebox[\linewidth]{\rule{\textwidth}{1pt}} 

The web is the most prominent application platform globally, thanks to its vast user base. Although it started explicitly as a site platform in the 90s, it evolved into an application platform over time as its potential as such was recognized and interactive web applications became a reality. So-called single-page applications (SPAs) represent the current mainstream approach for developing complex web applications. While SPAs provide a good experience for developers, they come with a cost of their own for the users due to the technical assumptions underneath. Disappearing frameworks question these technical assumptions and allow developers to address user needs better while retaining the benefits of the earlier approaches.

In this short book, I want to give you a quick overview of the latest developments in the field while explaining why I believe disappearing frameworks will inspire the shape of web development during the coming years. Many of the ideas are accessible already, and after reading this book, you will know where to look when evaluating new frameworks while being able to appreciate their level of innovation better. At the same time, you will see the current mainstream frameworks in a different light and understand their technical constraints better.

We start by delving into the history of web development to understand what has motivated the development of disappearing frameworks and why it is such an important topic. You could say that disappearing frameworks emerge from the pressures of both users and developers as the threshold for what is expected from a web application rises each year while developers are expected to deliver faster. It is within this intersection where technical innovation occurs as new tooling can give a higher baseline enabling developers to deliver more quickly and more robust applications for their users.

\mainmatter


\chapter{Brief history of web development}
\label{ch:brief-history}
It is good to understand the background in brief detail to put the history of web development into context. As early as 1965, two computers at MIT Lincoln Lab communicated using packet-switching technology \cite{zimmermann2022}. The work at MIT Lincoln lab provided a precursor for TCP/IP protocol\footnote{TCP/IP forms the backbone of the internet and specifies how communication should occur in computer networks.}, the backbone of the internet. ARPANET, the predecessor of the internet, was launched in 1969. The internet started in 1983 as ARPANET and Defense Data Network (DDN) moved to use the newly specified TCP/IP protocol \cite{maybaum1986defense,zimmermann2022}.\footnote{\href{https://www.amazon.com/How-Internet-Happened-Netscape-iPhone/dp/1631493078}{How the Internet Happened: From Netscape to the iPhone} and \href{https://yle.fi/aihe/a/20-10003269}{Halt and Catch Fire at Yle Areena} cover the story in greater detail.}

In 1992, \textbf{world wide web} (www) combined the ideas of hypertext and information retrieval through access systems in the seminal work of Berners-Lee from CERN \cite{bernerslee1992}\footnote{www was not the first system, but it is the one that became dominant. Gopher is a good example of a predecessor that faded away, and its history has been covered in detail at \cite{frana2004before}}. The web allows information retrieval and search through hyperlinks. At the same time, websites can consist of data indices and individual pages that may either exist on a file system or be completely virtual and generated on demand using a web server through a process called \textbf{server-side rendering} (SSR). A few years later, search engines relying on crawling the web\footnote{It is not a surprise the tools performing crawling the web are called spiders.} emerged to index the web and to make it easier to find information from it. \index{world wide web} \index{server-side rendering}

\section{Clients and servers}

The web has been built on top of the concepts of clients and servers. Generally, by clients, we refer to the users or, more specifically, to their machines. The machines, in turn, can be desktop computers, laptops, mobile phones, televisions, and even smart watches, as web browsers exist in the most unexpected places these days. Servers form the backbone of the Internet as they do most of the computational work. At the most superficial level, servers relay files to the client from a file system while complex servers communicate with other resources to compute responses to the client. Web techniques exist within this spectrum, and there are also different takes on how much logic should be on the client side. It is within this spectrum where disappearing frameworks appear later in this book.

\begin{figure*}
  \includegraphics{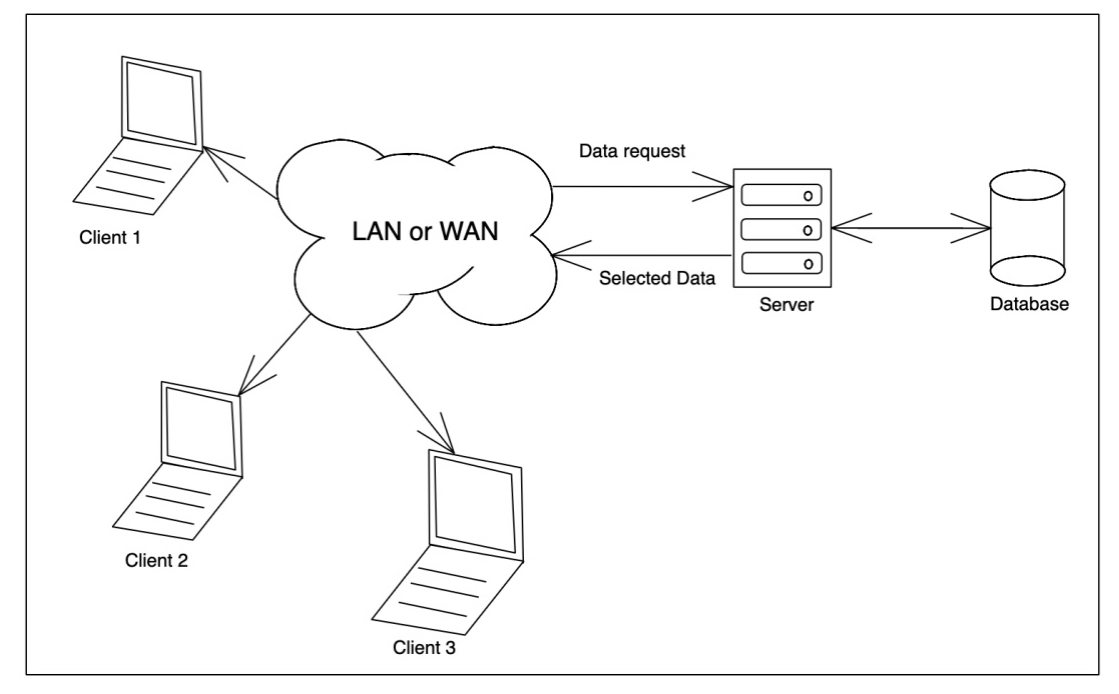}
  \caption{Example of client to server communication \citep{kumar2019}}
  \label{fig:kumar}
\end{figure*}

Figure \ref{fig:kumar} shows how clients connect to servers in a classic web architecture. The model has remained relatively stable, although more layers exist in the modern web. Also, peer-to-peer communication from client to client has become possible in certain cases\footnote{\href{https://www.scuttlebutt.nz/}{Scuttlebutt} is an example of a social network that has been designed to be decentralized out of the box.}.

\section{cgi-bin (1993) - an early attempt at describing servers}
\index{cgi-bin}

cgi-bin was an early attempt to describe web servers coming from National Center for Supercomputing Applications (NCSA) \cite{cgibin}, and the example below shows how to implement a server using cgi-bin:

\begin{lstlisting}[language=Perl, caption=cgi-bin example]
#!/usr/bin/env perl
=head1 DESCRIPTION

printenv - a CGI program printing its environment

=cut
print "Content-Type: text/plain\n\n";

for my $var ( sort keys %ENV ) {
    printf "%s=\"%s\"\n", $var, $ENV{$var};
}
\end{lstlisting}

The example gets to the point in the sense that it captures a client request and returns a response. To communicate better what the client wants, specific HTTP verbs\footnote{HTTP verbs capture the intent of a request, and depending on the verb, the way query parameters are passed may differ (i.e., GET vs. POST).} is used, and standards exist to define server responses as well. Regardless of the technical stack used, these standards remain constant, although there may be slightly different interpretations of HTTP verbs, for example, depending on the architectural style. The difference is visible when comparing RESTful approaches with GraphQL, for instance.

\section{PHP (1994) - an accidental programming language for developing large-scale web services}
\index{PHP}

PHP, originally Personal Home Page Tools, started as a small set of cgi-bin scripts written in C by Rasmus Lerdorf, and originally PHP covered a range of everyday web-related tasks (logging, tracking, access, server-side includes, etc.) people needed to perform at that time \cite{lerdorf2002}. It was only due to a large project that Rasmus had to implement for the University of Toronto that pushed PHP toward the direction of a language, as the script-based approach did not scale well enough. The trend was visible in PHP 2, and starting from PHP 3, where Rasmus collaborated with Zeev Suraski and Andi Gutmans, PHP took off as a technology, and the rest is history, given PHP became one of the most popular languages for web development \citep{lerdorf2002}.

\section{JavaScript - a language for scripting the web (1995)}
\index{JavaScript}

It is difficult to imagine the modern web without JavaScript\footnote{Due to historical reasons, JavaScript is a trademark of Oracle Corporation.} as it is the programming language used in the frontend, and occasionally even in the backend, making it a proper full-stack language. Initially, JavaScript was designed to add interactivity to webpages and to complement Java, hence the name. However, over the longer term, it grew as a complete language that can be used for any imaginable task \cite{wirfs2020javascript}. Consider the example below to get a quick idea of what JavaScript looks like:

\begin{lstlisting}[language=JavaScript, caption=JavaScript example]
function hello(name) {
  console.log(`hello ${name}!`);
}

hello("world");
\end{lstlisting}

The story of JavaScript is a complex yet interesting one. \citep{wirfs2020javascript} and \cite{vepsalainen2023ecmascript} cover the story of ECMAScript, the standard version of JavaScript, from a standardization point of view which helps you to understand why the language is the way it is, how it is continuously evolved further, and how to participate in the process.

It is easy to argue that JavaScript is the most essential programming language for web developers. Given it is flexible by definition, languages such as \href{https://www.typescriptlang.org/}{TypeScript} provide rigidity on top of it in the form of types. JavaScript may gain at least type hints one day, but even then, having a type system on top of authoring JavaScript can be beneficial. The trend is visible in modern runtimes\footnote{Given JavaScript is an interpreted language, it needs to be run through a specific runtime that can evaluate the code upon execution. Web browsers include runtimes of their own to evaluate JavaScript.}, such as \href{https://deno.com/}{Deno}, that support TypeScript out of the box.

\section{Document Object Model (DOM) (1998)}
\index{DOM}

JavaScript alone is not enough for developing web applications. ECMAScript specification doesn't cover its interaction with the browser on purpose \citep{wirfs2020javascript}. Instead, the Document Object Model (DOM) describes the browser interaction layer, which is a platform- and language-neutral interface that allows programs and scripts to dynamically access and update the content, structure, and style of documents \cite{wood1998}. Before DOM was specified, every browser, editor, and transformation engine handled things their way \cite{phillips1998}. With DOM, you can write scripts for different browsers and editors \citep{phillips1998}. Specifically, the DOM provides a programmatic API for interacting with the DOM structure, and when people speak of using the platform in the web context, they most likely mean the DOM.

Given the low-level nature of the DOM and occasional omissions, many abstractions have appeared on top of it. \href{https://jquery.com/}{jQuery} is perhaps one of the most famous of the abstractions as during its introduction in 2006, it normalized between browser APIs. jQuery gave a simple chaining API for everyday operations to make a webpage come alive with interactivity. Since then, the DOM API has caught up, and these days it provides a query API (\href{https://developer.mozilla.org/en-US/docs/Web/API/Document/querySelector}{querySelector}) comparable to jQuery to provide improved ergonomics for web developers. To give an example of DOM usage, consider the code sample below:

\begin{lstlisting}[language=JavaScript, caption=DOM example]
// Select all link elements
document.querySelectorAll("a");

// Select the first element with the class outline
document.querySelector(".outline");

// Select the first input with the attribute of name
// matching to login
document.querySelector("input[name='login']");
\end{lstlisting}

\section{CSS (1994)}
\index{CSS}

In 1994 as the web was gaining popularity, there was no way to style documents uniformly across different web browsers. For Tim Berners-Lee, the styling was not a priority, and multiple competing approaches appeared \cite{csstricksLookBack}. Demand from content authors led to the birth of the first draft of CSS by Håkon Wium Lie in October 1994 \cite{csstricksLookBack}. To ensure browser compliance, so-called \href{https://www.acidtests.org/}{Acid Tests} were developed starting from 1998 \cite{acid1}.

CSS can be characterized as an \textbf{aspect oriented programming} (AOP) language as it comes with selectors through which functionality, styling in this case, is attached to the pages. The approach is flexible as it is decoupled from content; styling is applied on top of it. The example below shows what CSS looks like:

\begin{lstlisting}[language=CSS, caption=CSS example]
.selector {
  margin: 2em;
  padding: 4em;
  color: red;
}
\end{lstlisting}

\section{Elements of a website/application}
\label{sec:elements}

\begin{marginfigure}
    \includegraphics{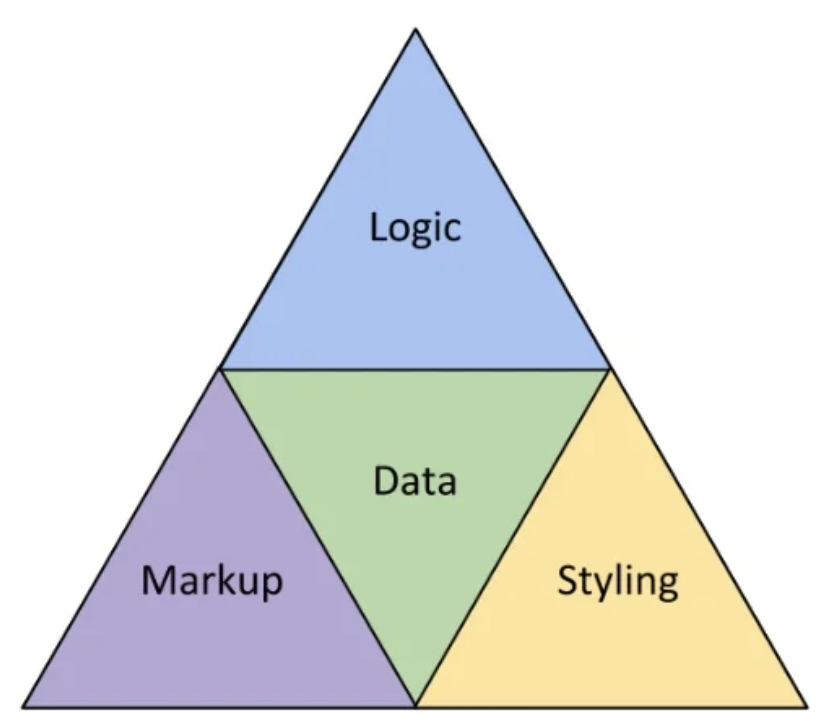}
    \caption{Elements of a web application}
    \label{fig:elements}
\end{marginfigure}

Regardless of the technical approach, each website and application consists of four distinct parts, as shown in Figure \ref{fig:elements}: data, markup, logic, and styling. By data, we refer to the information the site wants to show. The data is formatted using markup, and typically this operation is called templating, and the solutions doing this task are known as templating engines. Styling refers to how the markup is formatted to the user visually, and typically CSS language is used for the task as it has become the standard approach for the web. Finally, logic refers to the code executed to generate the response and potentially performed on the client side.

\section{Content delivery networks (CDNs) (1998)}
\index{CDN}

\begin{figure}
    \includegraphics[scale=0.305]{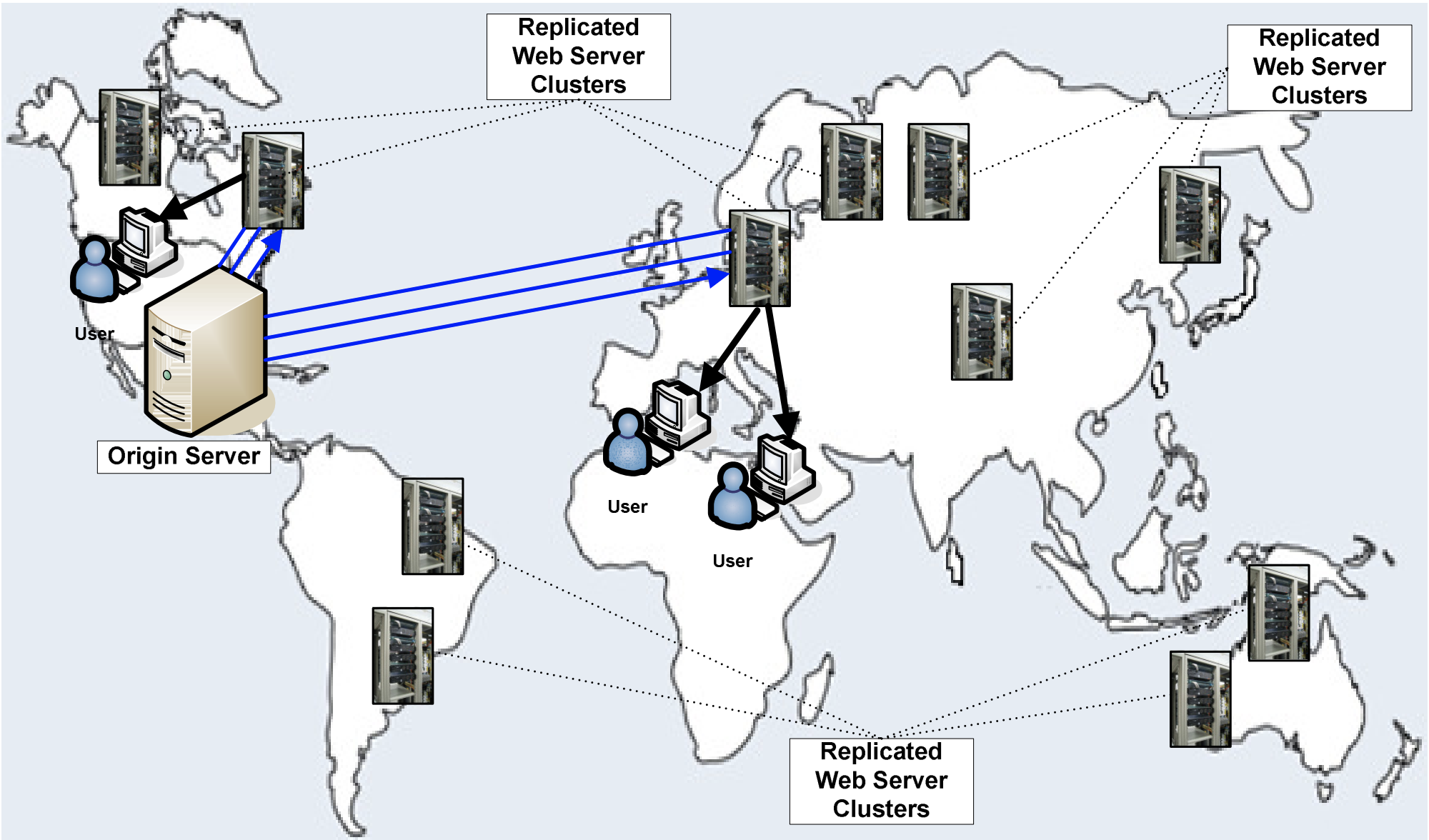}
    \caption{Content delivery network \citep{pathan2007taxonomy}}
    \label{fig:cdn}
\end{figure}

\textbf{Content delivery networks} (CDNs) represent a step forward in how static content can be delivered to clients by delivering content from a server close to a client through a distributed network \cite{nygren2010akamai}. These \textbf{points of presence} (PoP) replicate content and avoid problems serving it all from a single location or server. Beyond scalability, serving content from a PoP close to a client gives benefits in terms of network latency. In other words, the clients receive the content fast regardless of their location, depending on the density of the CDN network. Figure \ref{fig:cdn} illustrates how content is delivered through a CDN network to the clients.

CDNs were invented by Daniel Lewin, one of the founders of Akamai, in 1998 \cite{danielLewinCDN} and have since changed how content is delivered over the web. CDNs represent one of the building blocks in the modern web infrastructure, and so-called edge computing expands the usefulness of CDN networks to the next level.

\section{Edge computing}
\index{edge computing}

\begin{figure}
    \includegraphics{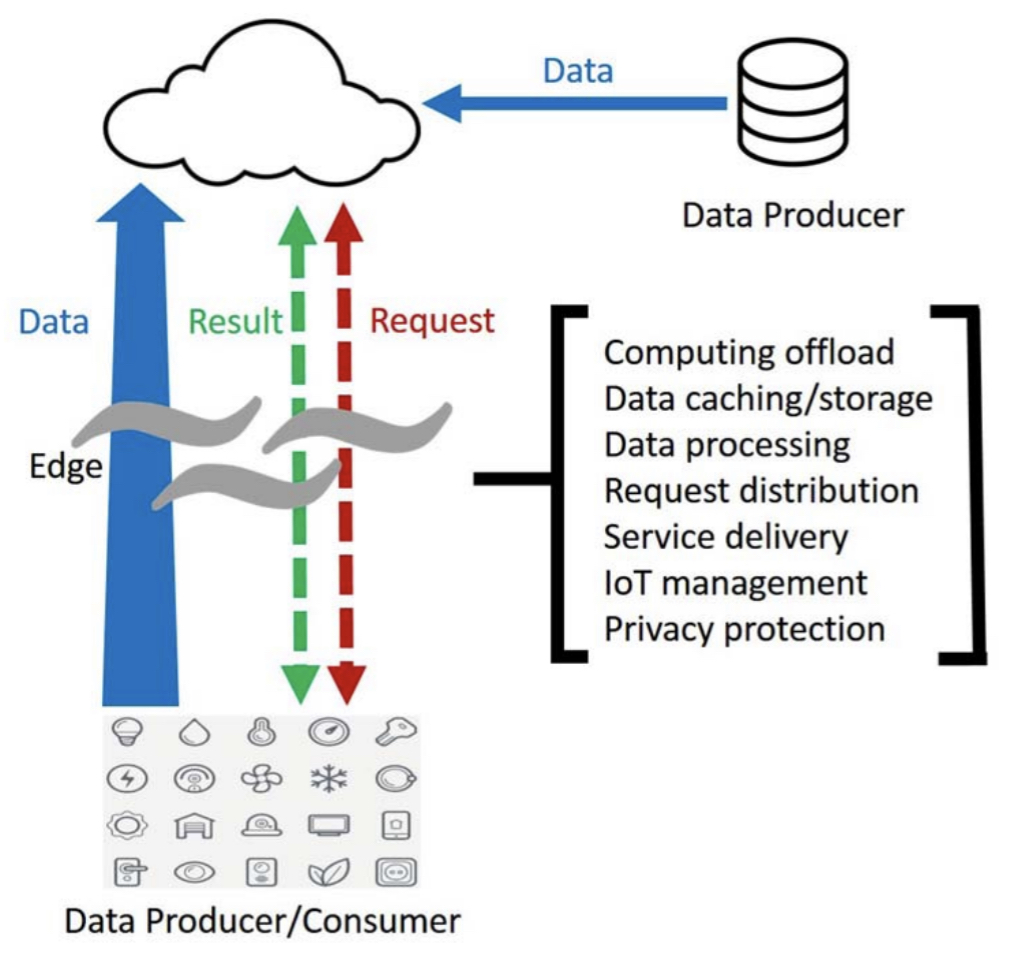}
    \caption{Edge computing \citep{weisong2016}}
    \label{fig:edge-computing}
\end{figure}

By \textbf{edge computing}, we mean computation happening on a server close to a client. Rather than referring only to the delivery of static content, an edge worker can perform logic and then return content based on that. An edge worker can also exist between a client and a server as middleware and perform some form of computation between them. An excellent example of this type of task is AB testing, where the edge worker would route the client to a server chosen by their test cohort. Edge computing can be used for tasks beyond this. The edge has become available to web developers only in the past few years by companies such as Amazon, Cloudflare, Deno, Netlify, and Vercel. Figure \ref{fig:edge-computing} shows edge computing flow and possible tasks that can be performed by using the infrastructure.

\pagebreak

\section{Conclusion}

Web as we know it is roughly thirty years old. During these thirty years, the web evolved significantly and morphed from a site platform to an application. Even from the beginning, there were building blocks of modern web applications although they were greatly refined over time. As time went by, so did the user expectations and the standards followed enabling developers to create even better user experiences on top of the web. Today the web is the largest application platform available, covering as much as roughly two-thirds of the global population \cite{statistaInternetSocial}.

\chapter{How did we arrive at disappearing frameworks}
\label{ch:what-are-disappearing-frameworks}
Disappearing frameworks represent the latest step in the evolution of web application development. As seen earlier in the history chapter at \ref{ch:brief-history}, the web evolved to become an application platform over time and it was never designed as such although web standards eventually enabled it to become an application platform. The evolution of the web has gone through distinct phases and the same applies specifically to web application development. To understand what disappearing frameworks are and why they matter for modern development, it is good to consider how we arrived at the current point from an application point of view.

\section{Progressive enhancement (2008)}
\index{progressive enhancement}

As discussed earlier in \ref{sec:elements}, the basic elements of any website or application are logic, data, markup, and styling. To create a minimal page, only markup is needed and the rest is optional while adding value to the markup. \textbf{Progressive enhancement} is an idea that builds on top of this realization by claiming that it makes sense put focus on content first while then building up the way it is presented through styling and finally applying client-side scripting, or logic as in our categorization \citep{alistapart2008}. Figure \ref{fig:progressive-enhancement} shows how the layers relate to each other:

\begin{figure}
    \centering
    \includegraphics[scale=0.5]{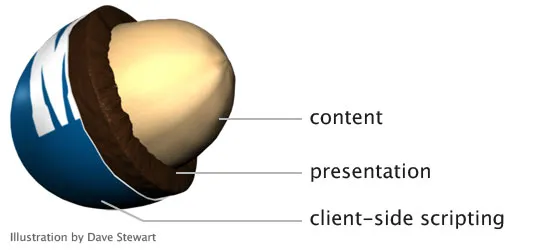}
    \caption{Progressive enhancement in a nutshell \citep{alistapart2008}}
    \label{fig:progressive-enhancement}
\end{figure}

Although the idea of progressive enhancement is fairly old as it was presented in 2008, there is still value in thinking in layers like this given the focus keeps our websites accessible and puts priority on content. The implication is that sites built using progressive enhancement work even if JavaScript or CSS have been disabled or are not available for some reason. CSS does not make much sense for a group of users due to a disability, for example, and then getting markup semantics right makes a big difference.

It can be argued that the problem that progressive enhancement solves does not make so much sense for web applications that rely heavily on logic. That said, it is still a good principle to keep in mind and it turns out it is an idea that is built in within the idea of disappearing frameworks by design.

\section{Sprinkles architecture (2006)}
\index{sprinkles architecture}
\index{jquery}

One good way to follow the principle of progressive enhancement is to apply JavaScript logic separately through selectors. I call this type of approach the \textbf{sprinkles architecture} as you sprinkle logic on top of your markup. \href{https://jquery.com/}{jQuery} (2006) is one of the early pioneers in the space and since then it has become immensely popular as close to 78\% of websites use it \cite{jQueryw3techsUsageStatistics}. To understand how jQuery works, consider the following example where we use a CSS selector and then apply a click handler on the matching elements:

\begin{lstlisting}[language=JavaScript, caption=jQuery example]
$('.selector').on('click', () => alert('hello'))
\end{lstlisting}

It is this type of chaining syntax that made jQuery famous and on top of simple syntax, one of the early benefits of using jQuery was that it normalized the differences between browser APIs meaning you as a developer did not have to worry about corner cases or specific browser issues and could instead focus on writing your code \cite{bibeault2015jquery}. The early claim was that jQuery could reduce twenty lines of code written against the web platform with a mere three \citep{bibeault2015jquery}.

The architecture pioneered by jQuery remains relevant to this day and it has been reinterpreted by a group of newer solutions, such as \href{https://alpinejs.dev/}{Alpine.js} (2019) and \href{https://sidewind.js.org/}{Sidewind} (2019). In these cases, the idea has been to move a part of the declarations to the markup level while allowing JavaScript to be used if needed. The cost here is that JavaScript is still needed but that is something unavoidable.

jQuery does not provide structural support for developers and for that reason whole frameworks like \href{https://backbonejs.org/}{Backbone.js} (2010) and \href{https://knockoutjs.com/}{Knockout.js} (2010) were built. They added the structure needed to capture for example data handling-related concerns although they came with complexity of their own. For this reason, React, Angular, Vue, and others were developed to further simplify and capture the key ideas in a simple-to-digest form.

\section{Multi-page applications (MPAs) (1992) and Single-page applications (SPAs) (2008)}
\index{multi-page applications}
\index{single-page applications}

The early web applications were most often \textbf{Multi-page applications} (MPAs), where the application state lives on the server, and each request from the client to the server loads a new page \cite{kaluvza2018comparison}. It was only after the introduction of AJAX\footnote{AJAX as in Asynchronous JavaScript} in 1999 that more complex programming models became possible \cite{flanagan1998}. So-called \textbf{Single-page applications} (SPAs) is the most famous model of these as it leveraged the capability to update the page visible to the user without a full refresh \citep{kaluvza2018comparison}. Starting from 2008, it became possible to control client-side routing as well \cite{vepsalainen2023rise}.

SPAs were designed to address the problems of the traditional, server-driven approach \cite{rich2021}. In the SPA approach, a JavaScript framework takes care of rendering and handling logic and as a result, more complex UIs become possible to implement. Tuomas Pöyry \cite{poyry2021} has compiled the table \ref{table:ssrspa} below comparing SSR with the SPA approach:

\begin{table}
    \begin{tabular}{ |p{7cm}|p{1cm}|p{1cm}| }
     \hline
     Feature & SSR & SPA\\
     \hline
     Pre-rendered HTML & Yes & No\\
     Updates without refresh & No & Yes\\
     Supports forms & Yes & Yes\\
     Offline support in modern browsers & No & Yes\\
     \hline
    \end{tabular}
    \label{table:ssrspa}
\end{table}

To further complicate matters, an application can be split over multiple entry points, or pages, each of which has a SPA of its own. MPAs try to strike a balance between traditional applications and SPAs. In an MPA, you lose the navigation benefit of an SPA but at the same time, you can use different technologies on the pages. That in turn may be useful when you are modernizing a legacy codebase and don't have the capability to perform all the needed work at once.

Although an improvement over traditional web applications, according to Rich Harris \cite{rich2021}, SPAs suffer from the following problems:

\begin{enumerate}
    \item \textbf{Performance} is decreased due to the bloat introduced by JavaScript frameworks
    \item \textbf{Tooling} related to the approach is complex, and therefore less resilient as it won't work without JavaScript
    \item \textbf{Accessibility} is a problem and there are subtle bugs as a result
\end{enumerate}

To understand the differences between the approaches better, consider the table \ref{table:mpavsspa} from \citep{vepsalainen2023rise} below:

\begin{table*}
    \begin{tabular}{|l|c|p{7cm}|}
     \hline
     Dimension & MPA & SPA\\
     \hline
     Relies on JavaScript & No & Yes \citep{kaluvza2018comparison,solovei2018difference}\\
     Initial cost of loading & Potentially low & High due to dependency on JavaScript \citep{solovei2018difference} \\
     Overall response time & Slower & Faster due to partial updates \citep{kaluvza2018comparison}\\
     Business logic & Coupled & Decoupled \citep{kaluvza2018comparison,solovei2018difference}\\
     Refresh on navigation & Yes & No \citep{solovei2018difference}\\
     Bandwidth usage & Higher & Lower due to only transaction-related data moving between the parties \citep{kaluvza2018comparison} \\
     Offline support & Not possible & Possible \citep{kaluvza2018comparison,solovei2018difference}\\
     Search Engine Optimization (SEO) & Excellent & Possible but difficult \citep{iskandar2020comparison,kaluvza2018comparison,solovei2018difference}\\
     Security & Understood & Practices still being established \citep{kaluvza2018comparison} \\
     Routing & At server & Duplicated in server and client \citep{solovei2018difference}, but modern frameworks, such as Next.js, mitigate the problem \\
     \hline
    \end{tabular}
    \label{table:mpavsspa}
\end{table*}

\section{User experience (UX) and developer experience (DX)}
\index{developer experience}
\index{user experience}

When considering practices like progressive enhancement or sprinkles architecture, it is good to keep in mind why we are using them. For this reason, we generally speak about \textbf{user experience} (UX) and developer experience (DX) and consider implications to them when evaluating options. By UX, we mean how the end-user of a website experiences the site. UX includes topics such as web performance, UI practices, and accessibility. DX gives another, developer-centric view that focuses on how tooling works and enhances the capabilities of a developer allowing them to develop faster for example. Another way to put it is that DX considers tools from a productivity point of view.

The earliest web development tools had a clear focus on UX as the target was to author websites and the authoring flow might have been purely text-based with slow feedback loops. Early graphical editors, such as Frontpage or Dreamweaver, gave a graphic way to design webpages while compromising on the resulting markup, therefore, emphasizing DX at the cost of UX. CMS systems, such as WordPress, went further by introducing more roles, such as content creators or designers, and allowing them to work with the system instead of pure code.

It can be argued that sometimes technical solutions go too far in one direction. For example, the recent trend of SPAs can be criticized for its DX focus while forgetting about UX. It is fast and easy to develop a SPA but how do you know it has a good UX? As a result, developers have to consider UX separately. Instead, it would be better to have a higher baseline for both, and that is one of the selling points of disappearing frameworks as they approach the problem from a new angle while addressing both UX and DX unlike solutions before.

\section{Why client-side performance matters}

\cite{dellaert1999tolerable} found in their classic study that waiting can affect evaluations of websites negatively although it necessarily does not have to be so if the waiting experience is well managed. Furthermore, \cite{an2018} found that as waiting time goes up, so does the so-called bounce rate. In other words, the longer users have to wait, the more likely they are to leave the site. \cite{stadnik2018impact} supports the finding in terms of a correlation between load time and e-commerce conversion rate. A slow site is worst of the both worlds as it has an increased bounce rate and a lower conversion than it would otherwise have. For this reason, it makes sense, especially for e-commerce-facing companies to put a definite focus on web performance as it has a clear payoff in terms of investment.

\begin{figure*}
    \includegraphics{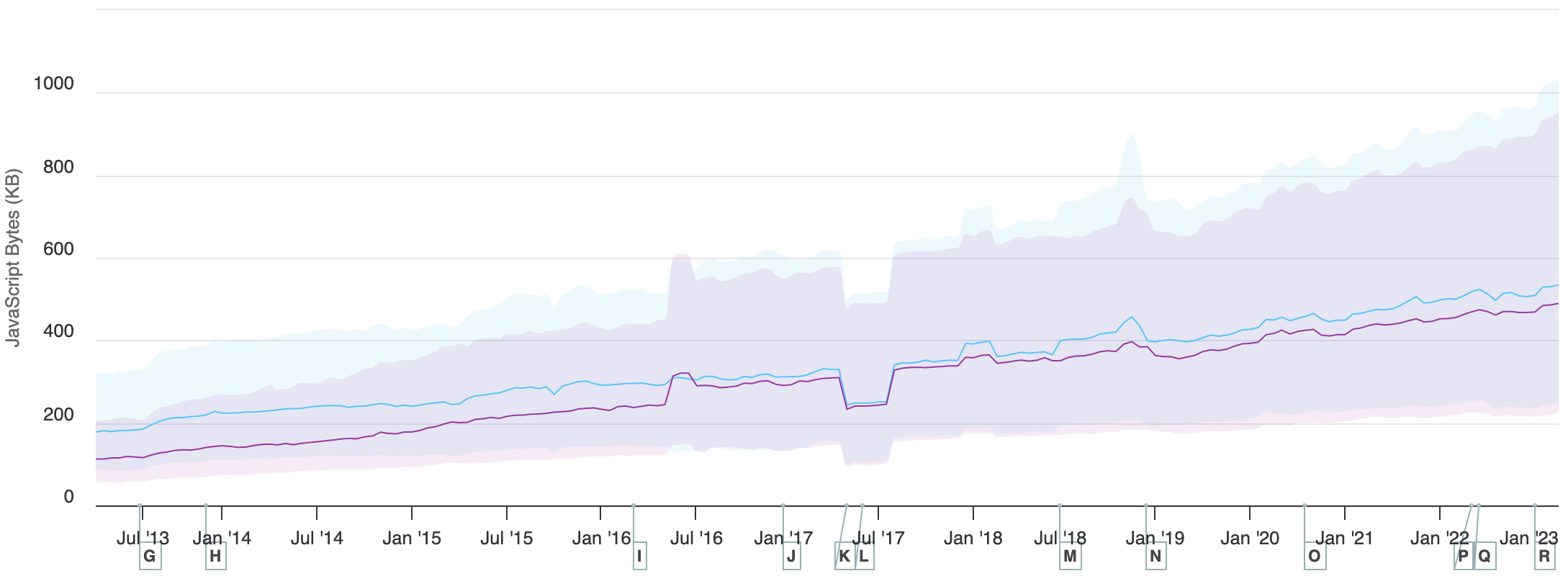}
    \caption{Amount of JavaScript on websites over time \citep{httparchiveHTTPArchive}}
    \label{fig:js-trend}
\end{figure*}

As shown by \cite{httparchiveHTTPArchive}, the amount of JavaScript shipped to the client has been steadily increasing over time as seen in Figure \ref{fig:js-trend}. Likely the increase has to do with the current, JavaScript-heavy development trends, and as shown by research, it is a trend that is costing companies money and users time not to mention environmental cost. It can be argued that it is time for a change in development practices and that is what disappearing frameworks are all about as they remedy the improvements made to DX over the past decade with the need to ship less to the client.

\section{Transitional Web Applications (2021)}
\index{transitional web applications}

To overcome the issues of the earlier approaches, Rich Harris proposed the concept of \textbf{transitional web applications} (TWA) in his groundbreaking talk \cite{rich2021} (2021). TWAs take ideas from both the traditional web and SPAs, and they have the following characteristics:

\begin{itemize}
    \item \textbf{Server-side rendering} is utilized to provide fast initial loading times
    \item \textbf{Resiliency} is achieved by allowing the applications to work without JavaScript by default
    \item \textbf{Consistent experience and accessibility} are built in by definition
\end{itemize}

In his analysis \cite{ryan2021}, Ryan Carniato further points out that while SPAs come with an excellent \textbf{Developer eXperience} (DX), they direct people to use a lot of JavaScript and ignore native browser APIs. Combining SPAs with SSR didn't change much due to performance concerns related to hydration.

\section{Disappearing frameworks}
\index{disappearing frameworks}

In contrast to the current mainstream JavaScript frameworks, the starting point for disappearing frameworks has to do with delivering the minimal amount, or even zero, of JavaScript to the client \cite{vepsalainen2023rise}. The background and techniques discussed in this chapter form the reasoning for why disappearing frameworks are needed and form a natural evolutionary direction for web development. The shift in focus has allowed new approaches to emerge and at the same time, the technical challenge is to retain the DX benefits of the current frameworks while not veering too far on a conceptual level. 

As highlighted in \citep{vepsalainen2023rise}, current frameworks each follow three fundamental principles: component orientation, templating, and hydration. Component orientation means that a user interface (UI) can be modeled through a component abstraction where components may be composed freely. Furthermore, components may capture specific portions of the UI. By templating, we refer to the markup used to define components and commonly this is some form of DSL\footnote{DSL as in domain-specific language.}, such as JSX\footnote{JSX was popularized by React, and although it was initially controversial, JSX has become accepted as a standard. JSX allows developers to mix JavaScript syntax they are familiar with HTML-kind syntax.}, which is mapped to HTML. Hydration is the process during which the client code is evaluated, and the UI is made interactive. In combination, these three form the core of the current mainstream JavaScript frameworks. In disappearing frameworks, especially the principle of hydration receives critical thought as it may not be needed after all.

\section{Islands architecture (2019)}
\index{islands architecture}

\begin{figure*}
    \includegraphics{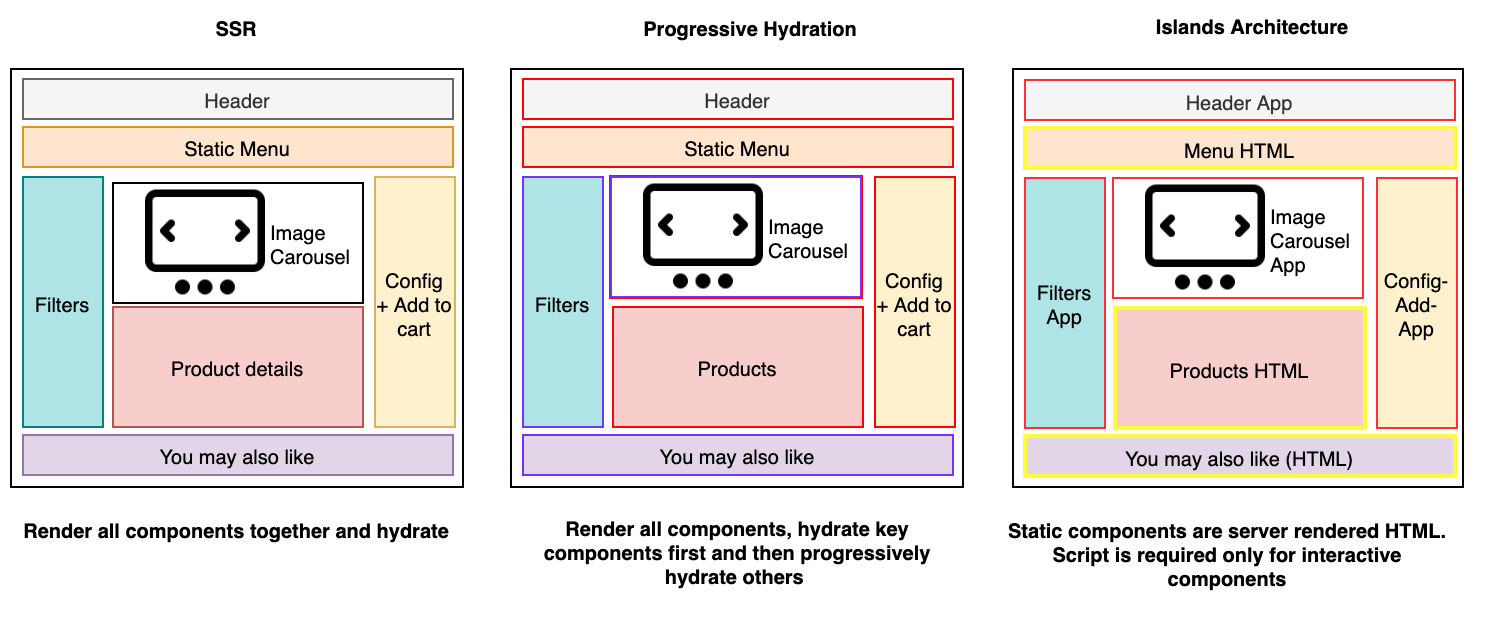}
    \caption{Islands architecture compared \citep{patterns2022}}
    \label{fig:islands}
\end{figure*}

Islands architecture can be considered as a stepping stone towards disappearing frameworks \cite{vepsalainen2023rise}. The idea of islands architecture is to imagine that a web page consists of slots that can be either static or dynamic by their nature. The dynamic sections of a page are what we consider islands in the approach and on top of this the architecture allows the developer to decide when the code related to an island is loaded by using specific loading strategies. Having this extra control through strategies allows developers to emphasize the most important functionality while loading secondary functionality later as it is needed for example. Figure \ref{fig:islands} shows how islands architecture compares with SSR and progressive hydration\footnote{In progressive hydration the key components are hydrated first and the rest later \cite{patterns2022}.}

Given islands architecture was formalized only in 2019, there is not much experience in using it yet \citep{jason2020}. Therefore the boundaries of the architecture are not fully understood yet, but at least it provides a compromise between fully dynamic and static sites thereby expanding the spectrum. At the moment, only several frameworks support islands out of the box\footnote{Consider checking out \href{https://astro.build/}{Astro}, \href{https://capri.build/}{Capri}, \href{https://fresh.deno.dev/}{Fresh}, \href{https://iles-docs.netlify.app/}{îles}, and \href{https://markojs.com/}{Marko} for example.} although the architecture can be implemented on top of any existing solutions and libraries, such as \href{https://github.com/11ty/is-land}{11ty/is-land}, make the task easier.

\section{Conclusion}

Disappearing frameworks form a new way to approach web application development by shifting focus to delivering less to the client and by being more thoughtful about it. Earlier technical solutions have pushed the state of web application development considerably while user expectations have increased. At the same time, mobile usage of the web has increased and the way we use the web has become more diverse while the size of websites keeps growing. Disappearing frameworks address the challenge by going back to the basics in some ways while picking up the improvements made by current mainstream JavaScript frameworks. In short, disappearing frameworks bridge the concepts of UX and DX within a single package and provide a higher baseline for developers for developing their applications.

\chapter{Examples of disappearing frameworks}
\label{ch:examples}
Although not a single new framework claims to be a disappearing one, several already implement the idea as we defined in the previous chapter. Given that the target of disappearing frameworks is loosely defined, there is plenty of room for technical innovation and approaches. The following example projects illustrate the variety of systems and also provide ideas on how you might proceed in implementing your own disappearing framework.

\section{Astro}
\index{Astro}

\href{https://astro.build/}{Astro} characterizes itself as the all-in-one web framework designed for speed. Another way to think about Astro is that it could be said to be a \textbf{metaframework}. In other words, to use Astro, you must bring your user interface framework, such as React, Svelte, or Vue, to accompany it. Due to its approach, you can also use multiple in the same project making Astro an interesting option for legacy projects.

\subsection{Astro islands}

Astro is the framework that popularized the concept of islands architecture, and islands are easy to use within it as illustrated below based on their documentation \cite{astroAstroIslands}:

\begin{lstlisting}[language=JavaScript, caption=Astro islands example - \textbf{src/pages/index.astro}]
---
// Example: Use a dynamic React component on the page.
import Counter from '../Counter.jsx';
---
<!-- This component is now interactive on the page! 
     The rest of your website remains static and zero JS. -->
<Counter client:load />
\end{lstlisting}

The example has two interesting points: file format and how the island is defined. Astro uses specific \textbf{.astro} files that contain what is called a headmatter and the template. In this case, the content is React code, and the component in question is imported within the headmatter. The headmatter would be the correct place for concerns such as data fetching and injecting content into the template context. In the example, the template uses the component imported through the headmatter and applies a specific loading strategy, \textbf{client:load}, to it. The page would render without the strategy, but Astro would not convert it to an island in that case. Consequently, the related logic would not work as Astro would render only static markup for the component instead.

\subsection{What makes Astro a disappearing framework}

Astro is an excellent example of a framework designed to work with mainstream user interface frameworks. Astro provides structure around them, allowing new, lighter ways to compose websites and applications. Astro implements islands architecture as a first-class concept and can be considered a disappearing framework by design.

It can be argued that there are limitations to its approach, but at the same time, it can be considered a bridge type of solution that improves the current situation. As we will see in the following examples, a disappearing framework can go even further than Astro, although compatibility with the existing user interface framework may have to be compromised.

\section{Fresh}
\index{Fresh}

\href{https://fresh.deno.dev/}{Fresh} claims to be the next-gen web framework built for speed, reliability, and simplicity. Fresh achieves these targets by rendering just-in-time on top of the edge while supporting island architecture out of the box. Like Astro, it does not ship JavaScript to the client by default. Furthermore, Fresh avoids a build step and configuration while supporting TypeScript out of the box. To achieve this all, Fresh has been built on top of Deno.

\subsection{What is Deno}
\index{Deno}

Fresh has been built on top of \href{https://deno.com}{Deno}, a new JavaScript runtime developed by the original author of \href{https://nodejs.org/en}{Node.js}, Ryan Dahl. In Deno, he sought to fix the mistakes he made in the design of Node.js. Incidentally, the development of Deno has inspired Node.js to improve and adopt features, such as native \textit{fetch} or security defaults, pioneered by Deno. When comparing Deno to Node.js, Deno can be considered a toolkit with a robust standard library. In other words, Deno comes with everything you need for building servers out of the box, including tooling for formatting your code, testing, running tasks, and whatnot. In contrast, within the Node.js ecosystem, you, as a developer, are forced to put together your development setup from the tooling provided by the community.

\subsection{What makes Fresh a disappearing framework}

Like Astro, Fresh has been designed to ship minimal JavaScript. Compared to Astro, it is more limited as Fresh has been built around Preact user interface framework. Preact is a light derivative of React that comes with a compatible API making it a good fit for Fresh. In other words, Fresh does not give any options beyond Preact to develop your user interface, limiting its usefulness in legacy situations. It also comes with a dependency on Deno which you may not prefer in your environment. More specifically, Fresh has been designed to be deployed on top of \href{https://deno.com/deploy}{Deno Deploy} edge platform, which is somewhat limiting as Astro does not have a similar limitation.

\section{Marko}
\index{Marko}

\href{https://markojs.com/}{Marko} from eBay combines server and client-side rendering techniques while streaming content. Marko can tell which components to render on the client and which solely on the server through its compiler-based approach.

\subsection{Marko DSL}

Marko relies on a DSL that is used to describe pages, as shown in the example below adapted from the Marko landing page:

\begin{lstlisting}[language=HTML5, caption=Marko example - \textbf{src/routes/\_index/+page.marko}]
<!doctype html>
<html>
<head>
    <title>Hello Marko</title>
</head>
<body>
    <h1>My favorite colors</h1>
    <ul>
        <for|color| of=["red", "green", "blue"]>
            <li style=`color:${color}`>
                ${color.toUpperCase()}
            </li>
        </for>
    </ul>
    <shared-footer/>
</body>
</html>
\end{lstlisting}

To use Marko, you would have to learn their custom templating syntax. As with Astro and Fresh, a component abstraction is available; in the example, we are pointing to a \textbf{shared-footer} component.

\pagebreak

\subsection{What makes Marko a disappearing framework}

Given its e-commerce background and focus, Marko heavily focuses on performance. Marko achieves this through its compiler-based approach and custom DSL. Compared to Astro, it goes a notch further while compromising compatibility with mainstream user interface frameworks, although that is likely an acceptable compromise given the project's focus.

\section{Qwik}
\index{Qwik}

To quote \href{https://qwik.builder.io/}{Qwik} tagline, the project focuses on developing instantly interactive web applications without effort. Qwik includes a compiler, which controls both the front and backend, although it is possible to emit a completely static build for static use cases. Unlike Marko, however, Qwik relies on JSX and provides compatibility with React components for legacy projects.

Qwik's biggest attraction is the way it handles interactivity and avoids the cost of hydration by implementing the idea of resumability. Qwik can generate code loaded on a granular level through its code-splitting algorithm.

\subsection{Code splitting in Qwik}
\index{code splitting}

In the current generation of tools, like React, developers must manually take care of code splitting. By code splitting, we mean defining specific split points using the standard \href{https://developer.mozilla.org/en-US/docs/Web/JavaScript/Reference/Operators/import}{import()} syntax. The syntax tells the browser and bundlers, assuming a bundler\footnote{By a bundler, we mean a tool that analyzes project source code and emits code bundles that work in the browser. In other words, bundlers transform code from a developer-friendly one to a web browser-friendly format.} is used, and the related module should be loaded only when the code triggers initially.

The technique helps load heavy, rarely used components, for example, and the general idea is to defer the cost of loading by pushing it to the future. Sometimes, the cost will never occur as the user won't access the functionality, thereby saving bandwidth. The problem is that although code splitting is possible in the current generation of tools, it requires manual effort, and code splitting boundaries are not as granular as they could be.

In Qwik, the compiler handles code splitting automatically as it detects any \$ within the code and creates a split point there. When the user triggers the related code path, Qwik loads and executes the code, thereby deferring effort. To give further control to the developer over when the loading happens, Qwik exposes a \href{https://developer.mozilla.org/en-US/docs/Web/API/Service_Worker_API}{Service Worker} that can be used to define a more eager strategy if needed.

\pagebreak

The example below, adapted from Qwik documentation, shows how code splitting and state management work in Qwik:

\begin{lstlisting}[language=JavaScript, caption=Qwik example - \textbf{src/pages/index.tsx}]
import { component$, useStore } from '@builder.io/qwik';

export default component$(() => {
  const store = useStore({ count: 0 });

  return (
    <main>
      <p>Count: {store.count}</p>
      <p>
        <button onClick$={() => store.count++}>Click</button>
      </p>
    </main>
  );
});
\end{lstlisting}

\subsection{Resumability – a way to avoid hydration}
\index{resumability}

Unlike many other solutions available, Qwik avoids the cost of hydration. A conventional user interface, like React, has to execute the code on the client side to make the user interface interactive. Qwik uses a different approach called resumability. The idea is that the Qwik compiler calculates what it can on the server side and then passes the initial state to the client while converting the \$ portions of the code to a form understood by Qwik runtime (a couple of kB). Then at the client side, the runtime loads necessary interactivity on demand as described above.

\subsection{What makes Qwik a disappearing framework}

Qwik focuses on disappearing from the client, and through its approach, it tries to minimize the client cost related to using the framework. Compared to Astro, Qwik goes further because its compiler is doing more work, and using Qwik results in more granular loading behavior at the client. It is good to remember that the developer can still adjust the behavior as needed.

Qwik can be considered a full-stack framework as it allows developers to build both client and server-side logic, and it may even exist in the same files of the codebase, thereby improving cohesion. However, it can be argued the approach may leak unwanted information to the client.

Qwik follows the principle of progressive enhancement, and sites created using Qwik work without JavaScript enabled as necessary fallbacks are generated automatically as illustrated by its \href{https://qwik.builder.io/docs/action/}{form handling through actions}.

\section{Conclusion}

The example frameworks discussed in this chapter illustrate the benefits of the disappearing approach. By shifting the viewpoint, they provide developers with a higher baseline for building their applications in a user-friendly way. The solutions discussed take care not to lose the DX benefits gained through the era of SPAs and build on top of the tradition while merging it with the best ideas from the early web.

Interestingly enough, all of the solutions discussed have been built with edge platforms in mind and can easily be deployed on such. While Fresh is limited to be run on Deno Deploy, the rest of the frameworks discussed provide more options, and especially Astro and Qwik have support for many platforms out of the box.

\chapter{Where to go from here}
\label{ch:where-to-go-form-here}

For a developer, the question is can I start leveraging these tools already in my work? The answer is that it depends on the type of the project. Out of the projects discussed, Astro is the easiest one to adopt as it provides support for many of the mainstream user interface frameworks while giving clear benefits on top of them through its implementation of the islands architecture. In terms of design, Astro is not as aggressive as a solution like Qwik, but it is already a step in a good direction. A tool like Qwik is a bigger leap although for a React developer, it is comfortable to adopt in a new project. Therefore Qwik should be good for prototyping and smaller standalone projects although by design it should scale well even to extremely large projects as its client-side loading behavior has been heavily optimized.

For a tool author, the question is, how should I build my new framework? Based on what we have seen in this booklet, several trends come together in disappearing frameworks as they seem to combine the recent developments occurring in the edge computing space while leveraging the power of compilers to generate code that is optimal for both the server and the client. The cost is that a compiler provides yet another layer of abstraction and therefore may obscure from the developer what is happening underneath making the solution a black box to debug. Therefore the challenge is to retain the benefits of compilers while taking care to keep debuggability in mind.

Disappearing frameworks address clear demand and although it is still early days with the new breed of frameworks, they already provide tangible benefits for their users. It is likely the frameworks will evolve further and lead to so far unforeseen benefits perhaps in the form of new development practices and techniques. The starting point for disappearing frameworks is enticing as it gives a clear target for tool developers while giving benefits to all parties. Older practices, such as progressive enhancement, combined with the latest technical development come together in disappearing frameworks in a new way making a better web for all.

\backmatter

\bibliography{references}
\bibliographystyle{plainnat}

\printindex

\end{document}